\documentclass[fleqn,usenatbib]{mnras}

\usepackage{newtxtext,newtxmath}

\usepackage[T1]{fontenc}
\usepackage{soul}
\DeclareRobustCommand{\VAN}[3]{#2}
\let\VANthebibliography\thebibliography
\def\thebibliography{\DeclareRobustCommand{\VAN}[3]{##3}\VANthebibliography}

\usepackage{graphicx}	
\usepackage{amsmath}	

\title[Warp and flare of the Galactic disc]{Warp and flare of the old Galactic disc as traced by the red clump stars}

\author[Namita Uppal et al.]{
Namita Uppal,$^{1,2}$\thanks{E-mail: namita@prl.res.in}
Shashikiran Ganesh,$^{1}$
Mathias Schultheis$^{3}$
\\
$^{1}$Astronomy and Astrophysics division, Physical Research Laboratory, Ahmedabad, India\\
$^{2}$Department of Physics, Indian Institute of Technology, Gandhinagar, India\\
$^{3}$Université Côte d’Azur, Observatoire de la Côte d’Azur, CNRS, Laboratoire Lagrange, Bd de l’Observatoire, CS 34229, 06304 Nice cedex 4, France
}

\begin{document}
\label{firstpage}
\pagerange{\pageref{firstpage}--\pageref{lastpage}}
\maketitle

\begin{abstract}
Our study aims to investigate the outer disc structure of the Milky Way galaxy using the red clump (RC) stars. We analysed the distribution of the largest sample of RC stars to date, homogeneously covering the entire Galactic plane in the range of $40^\circ \le \ell \le 340^\circ$ and $-10^\circ \le b \le +10^\circ$. This sample allows us to model the RC star distribution in the Galactic disc to better constrain the properties of the flare and warp of the Galaxy. Our results show that the scale length of the old stellar disc weakly depends on azimuth, with an average value of $1.95 \pm0.26$ kpc. On the other hand, a significant disc flaring is detected, where the scale height of the disc increases from 0.38 kpc in the solar neighbourhood to $\sim 2.2$ kpc at R $\approx 15$ kpc. The flare exhibits a slight asymmetry, with $\sim 1$ kpc more scale height below the Galactic plane as compared to the Northern flare. We also confirm the warping of the outer disc, which can be modelled with $Z_w = (0.0057 \pm 0.0050)~ [R-(7358 \pm 368) (pc)]^{1.40 \pm 0.09} \sin(\phi - (-2^\circ.03 \pm 0^\circ .18))$. Our analysis reveals a noticeable north-south asymmetry in the warp, with a greater amplitude observed in the southern direction compared to the northern. Comparing our findings with younger tracers from the literature, we observe an age dependency of both the flare and warp.  An increase in flare strength with age suggests the secular evolution of the disc as the preferred mechanism for forming the flare. The increase of the maximum warp amplitude with age indicates that the warp dynamics could be the possible cause of the variation in the warp properties with age.
\end{abstract}



\section{Introduction}\label{Sec:1}
The Milky Way is a disc spiral galaxy having many substructures. The disc is the most massive and prominent component of our Galaxy that comprises approximately 75\% of all the Galactic stars \citep{Rix2013}. However, determining its morphology has been challenging due to two primary factors: Our vantage point inside the Galaxy, which limits our ability to observe the complete structure and second, the extinction of starlight by the dust present along the line of sight varying with direction.

 Technological advancements and the availability of large-scale sky surveys have allowed us to use star count methods to better understand the structure of our Galaxy \citep{Bahcall1986, Paul1993, Majewski1993,Bland2016}. Current understanding suggests that the number density of stars decreases exponentially as we move away from the Galactic centre both radially and vertically \citep[][etc.]{Gilmore1983, Bovy2012, Bland2016, Liu2017}. While we have gained significant knowledge about our Galaxy during the era of deep space surveys, many aspects still require further exploration. Most studies have focused on the structure of the Galaxy within a few kpc around the Sun. However, our understanding of the outer regions of the stellar disc is limited due to the low density of stars at large distances. The outer Galactic disc has many exciting features, such as warp and flare. Understanding the dominant processes in this part of our Galaxy is crucial for constructing galaxy formation and evolution models.

Warp is the large-scale distortion of the outer disc of the Galaxy with respect to the inner disc. Disc warps are common asymmetrical features present in the many disc galaxies \citep{sanchez1990, Reshetnikov1998, sanchez2003}, displaying various shapes like S, U, or L \citep{Kim2014}. The detection of warp in the Milky Way disc originated from 21 cm observations of neutral hydrogen (HI) \citep{Kerr1957, Oort1958} and subsequent confirmations were made by \citet{Weaver1974, Nakanishi2003, Levine2006a}. Since then, the warp has been studied extensively not only in neutral hydrogen but also using dust \citep{Drimmel2000, Marshall2006}, molecular cloud \citep{Grabelsky1987, May1997} and various stellar tracers and their kinematics \citep[e.g.,][]{lopez2002a, Momany2006, Reyle2009, warpAmores2017,Romero2019, Skowron2019a,poggio2020Nat, Li2019OB, Wang2020, chrobakova2021, Lemasle2022}. The disc of our Galaxy is curved upwards towards the north on one side and downwards on the south, depicting an S-shape warp. Despite being discovered 60 years ago and extensively studied in recent years, the exact shape and mechanism behind its formation are not well-constrained. Several theories have been proposed to explain the origin or the formation of the warp in disc galaxies. The theoretical models can be broadly divided into two classes. One possibility suggests that warps are the result of gravitational interactions, such as the interaction of satellite galaxies \citep{Burke1957,Kim2014, warpBailin2003, Weinberg2006, warpBailinJermy2003} or a misaligned dark matter halo,  galaxy mergers, which can perturb the disc and induce deformation. The second possibility involves the influence of non-gravitational effects like the accretion of intergalactic gas \citep{Ostriker1989,Quinn1992,lopez2002Mech} or interactions with intergalactic magnetic fields \citep{Battaner1990, warpBattaner1998} as a possible explanation of disc warping.
 The ubiquity of warps in disc galaxies and in all the tracers in the Milky Way suggests that they are either repeatedly regenerated or long-lived phenomena \citep{Sellwood2013}. 
\citet{Lopez2014}, analysed the star counts in 5-16 kpc range and reported that the warp is a long-lived feature. Nevertheless, the study led by \citet{Poggio2017, poggio2020Nat} using the OB and giant stars contradicts the steady evolution model of warp formation and suggests that the warp is a transient response of a disc to an outside interaction.
\citet{Nurhidayat2022} studied the warp using the RGB stars from Gaia DR2 data, also favours the non-static and dynamically evolving warp formation model.
Several lines of evidence suggest that the parameters of warp change with the average age of the tracers being used \citep{Drimmel2000, warpAmores2017,wang2018, Romero2019, Chrobakova2020, Wang2020, warpLi2023}. Age dependency has also been observed in the warp of external galaxies \citep[e.g.,][]{Radburn2014}. The difference in the warp amplitude with the age of the tracers implies that the warp is a long-lived but non-steady feature. 
On the other hand, the studies led by \citet{Poggio2018} and \citet{Cheng2020} 
did not find any age dependency on the structure. The age dependency on the warp needs further investigation to understand its evolution mechanisms.

In addition to the warp, another discernible feature observed in the outer Galaxy is the flaring of the disc. The flare is an increase in the scale height of the Galactic disc with a Galactocentric radius. It has been detected in various tracers including gaseous \citep[HI;][]{Nakanishi_HI_2003, Levine2006a}, molecular clouds \citep[CO;][]{Grabelsky1987, May1997}, dust \citep{Drimmel2001}, and the stellar components such as Cepheids \citep{Feast2014, Chakrabarti2015}, OB stars \citep{Li2019OB, wang2018, Yu2021}, older population \citep{Alard2000, lopez2002a, Momany2006, Reyle2009,Yu2021b}, SEGUE F and G stars \citet{lopez2014flare}, pulsars \citep{Yusifov2004}, and recently in supergiant and whole Gaia EDR3 data \citep{Chrobakova2022}. The presence of flare in all the tracers, including gas, dust and stellar components, suggests that it 
is not a transient but a persistent feature of the Milky Way disc. Despite extensive studies of disc flare in the last decade,  there are still ongoing debates regarding the flare parameters, its structure and origin.
 
 Most of the earlier work on warp and flare was based on the young stellar populations. However, there are a few efforts to trace the older population, either by using all the NIR/FIR stars \citep[e.g.,][, etc]{Drimmel2001,lopez2002a, Momany2006, Reyle2009} or RC / red giant branch stars in a limited region of the sky \citep[][]{Li2020RC, Wang2020, Yu2021,warpLi2023}. In this paper, we use the largest sample of RC stars ($\sim 8.8$ million), extracted in \citet{Uppal2023} (paper 1 hereafter), from the Two Micron All Sky Survey \citep[2MASS,][]{Skrutskie2006} colour-magnitude diagram with the aim to investigate the structure of the outer disc, especially the warp and flare.
 
 This paper is organized as follows. Section \ref{Sec:2} briefly describes the RC sample used in this study and the methodology to estimate the stellar densities. The results obtained from the radial and vertical distribution of the RC star sample are presented in Section \ref{sec:3}. This section also summarizes various analyses and results based on the modelling of flare and warp traced by RC stars.  Section \ref{sec:4} includes a detailed discussion of the age dependency of the flare and warp parameters by comparing the results obtained in this study with the other populations discussed in the literature. We summarize this work with the conclusion in Section \ref{sec:5}.

\section{Sample description \& methodology }\label{Sec:2}

RC stars are low-mass, core helium-burning stars having nearly constant luminosity and a small range in temperature during their phase of evolution \citep{Girardi2016}. This property makes them a good distance indicator and, hence, an excellent tracer to map the structure of our galaxy. We used a sample of $\sim 8.8$ million RC stars from paper 1, selected systematically from $1^\circ \times 1^\circ$ 2MASS colour-magnitude diagrams centred at $(\ell,b)$, covering the region bounded by $40^\circ \le \ell \le 320^\circ$ and $-10^\circ \le b \le 10^\circ$. In the sample, the RC stars were extracted by tracing the RC locus, discernible as the second peak in the density histograms of $J-K_s$ colours in J-bins. We imposed a limiting magnitude of J = 14.5 mag and required photometric uncertainties to be less than 0.1 mag in all three bandpass filters. The extent of colour dispersion varies across different lines of sight and was quantified as the $1\sigma$ colour spread around the central locus within each bin.
The contamination by foreground stars in the sample was removed using the high-accuracy astrometric data from Gaia Early Data Release 3 (EDR3). The details of the RC sample preparation and the validation of the catalogue are described in Paper 1.  2MASS data, even after twenty years from its data release, still plays an important role in selecting older populations at deeper distances as compared to Gaia which is confined to $\sim$ 5 kpc from the Sun. The distance and extinction of all the extracted RC stars were calculated from the distance modulus by considering the absolute magnitude and intrinsic colour of K2-type giants, which are representative of RC stars. The uncertainties in the distance measurements propagated from uncertainties in the absolute magnitude, intrinsic colour, and extinction are less than 10\%.
The sample selection was performed in Heliocentric coordinates. The extracted distance information, along with the position of stars, is converted to right-handed Galactocentric spherical coordinates using the following conversion:
\begin{equation}
    X = r\cos{\ell}\cos{b} - R_\odot
\end{equation}
\begin{equation}
    Y = r\sin{\ell}\cos{b}
\end{equation}
\begin{equation}
    Z = r\sin{b}
\end{equation}
Where r is the distance of the star from the Sun and $\ell$, $b$ are the Galactic longitude and latitude, respectively. The distance from the Galactic centre to the Sun is denoted by $R_\odot$ = 8.34 kpc \citep{Reid2014}. 

\subsection{Stellar density}\label{sec:2.1}
The sample utilized in this study provides information on the number of stars. To convert these star counts into density, a fundamental equation of stellar statistics described in equation \ref{eq:4} is employed, as outlined by \citep{Bahcall1986}. This equation allows the calculation of stellar density (D, stars pc$^{-3}$) within a specific solid angle $w = d\ell \times db$ from star counts (N).
\begin{equation}\label{eq:4}
    N(m) = w \int_{0}^{\infty} r^2 D(r) \phi(m) \,dr 
\end{equation}
Here, $N(m)$ is the number of stars per unit solid angle in the interval $m$ to $m+dm$, and $\phi(m)$ is the luminosity function which can be replaced with the delta since we are only considering K2-type giants. We consider the absolute magnitude of the RC stars; $M_J = -0.945$ mag \citep{Ruiz2018} in our analysis.  The density $D(r)$ at a distance $r$ from the Sun is calculated by dividing the sample of RC stars into Galactic longitude $\ell$, Galactic latitude $b$ and apparent magnitude $m$ with a bin size of $5^\circ$, $2^\circ$ and 0.2 mag, respectively using the equation \ref{eq:5} obtained by inverting the equation \ref{eq:4}.

\begin{equation}\label{eq:5}
    D(r) = \frac{N(m)\delta m}{w r^2 dr}
\end{equation}
The obtained mean density is converted into the Galactocentric cylindrical system using the following equations.

\begin{equation}\label{eq:6}
   R = \sqrt{R_\odot^2 +(r \cos{b})^2 - 2 r R_{\odot}\cos{b}\cos{l}}
\end{equation}

\begin{equation}\label{eq:7}
\phi = tan^{-1}\left(\frac{-Y}{-X}\right)
\end{equation}

\begin{equation}
    Z = r\sin{b}
\end{equation}
Where R is the radial distance from the Galactic centre, and Z is the disc height above or below the Galactic plane. The angle $\phi$ is the azimuth angle, and $\phi = 0^\circ$ points from the Galactic centre to the Sun, increasing counterclockwise. 

\section{Analysis and results}\label{sec:3}
We model the Galactic disc with RC stars by assuming a smooth density distribution model to infer the RC star counts in the Galactic disc. The stellar density in the disc is known to be decreasing exponentially with Galactocentric radius, R and height, Z above or below the Galactic plane \citep{Bland2016}. Therefore, we
started with a simple exponential model and then determined the effect of the flare and warp by adding the correction terms in the model. 
The density distribution is given by the equation \ref{eq:9}. 

\begin{equation}\label{eq:9}
    \rho(R,Z) = \rho_{\odot}~exp\left(-\frac{R-R_\odot}{H}\right)~exp\left(-\frac{|Z|}{h_Z} \right )
\end{equation}

Where $H$ and $h_Z$ are the scale length and scale height of the Galactic disc, $ \rho_\odot$ denotes the RC density in the Solar neighbourhood. In our analysis, we are only dealing with low latitude fields ($-10^\circ \le b \le 10^\circ$) where the contribution of the thick disc will be very small, so following \citet{lopez2002a}, we are using a single exponential function to model the disc. We call the disc an `Old disc' depicting the average structure traced by the intermediate-older population. 

\subsection{Radial distribution and scale length}\label{sec:3.1}
The density of RC stars falls exponentially as we move away from the Galactic centre. To simplify equation \ref{eq:9}, we are considering the distribution of stars only in the Galactic plane, i.e., for $b = 0^\circ$. For the Galactic plane $Z \approx 0$ and the equation \ref{eq:9} is then reduced to the following: 

\begin{equation}\label{eq:10}
     \rho(R,Z = 0) = \rho_{\odot}~exp\left(-\frac{R-R_\odot}{H}\right)
\end{equation}

\begin{figure}
\includegraphics[width=\columnwidth]{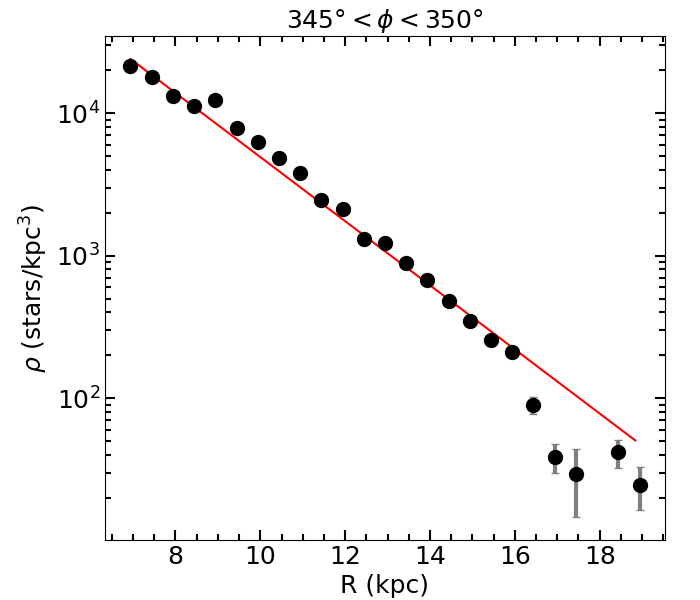}
\caption{Space density of RC stars in $\phi = [345^\circ, 350^\circ]$ as a function of Galactocentric mean distance in black points. The red line represents the best-fitted model with a solar neighbourhood density of $(1.16 \pm 0.04) \times 10^{-5}$ stars/pc$^3$ and scale length, H = $1.93 \pm 0.05$ kpc.  \label{fig:Fig1}}
\end{figure}

\begin{figure}
\includegraphics[width=\columnwidth]{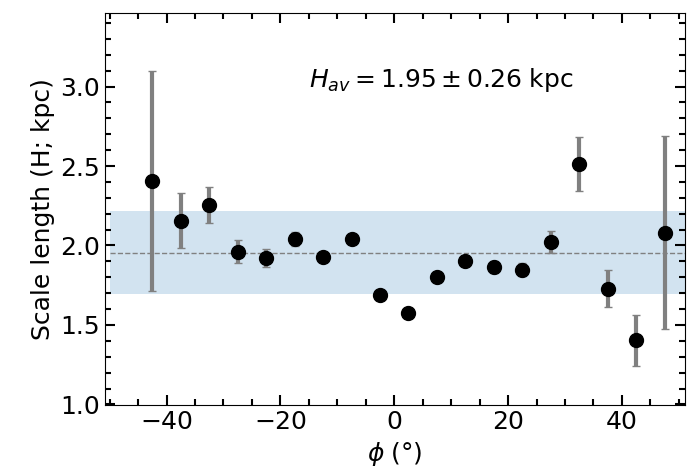}
\caption{Variation of scale length (H) as a function of azimuth ($\phi$). The dashed grey line represents the average scale length with $1\sigma$ dispersion highlighted as the shaded region. \label{fig:Fig2}}
\end{figure}

Here, we utilized $1^\circ$ bins in $b$, centred at $b = 0^\circ$ and $5^\circ$ bins in longitude, $\ell$ for a range $40^\circ \le \ell \le 340^\circ$ to calculate the stellar density following the procedure outlined in section \ref{sec:2.1}. The RC stellar density obtained in cylindrical coordinates is analysed as a function of Galactocentric radius in bins of $\phi$ with a bin size of $5^\circ$. One such distribution is illustrated in Figure \ref{fig:Fig1} for $345^\circ \le \phi \le 350^\circ$. The figure shows the decrease in stellar density (black points) as a function of the Galactocentric radius by taking 500 pc bins of R. To determine the solar neighbourhood density and scale length of the disc, we fitted equation \ref{eq:10} on the RC density in each $\phi-$bin.  
For the fit, we used `\textit{curve\_fit}'; function of \textit{SciPy} package from Python, which uses non-linear least squares to fit the function to the data. 
The best-fit model, shown by a red line in Figure \ref{fig:Fig1}, resulted in the scale length H = $1.93 \pm 0.05$ kpc and RC density in the solar neighbourhood of $\rho_\odot =(1.16 \pm 0.04) \times 10^{-5}$ stars/pc$^3$. The scale length obtained in different $\phi-$bins is plotted in Figure \ref{fig:Fig2}.  We consider the range in $|\phi| < 50^\circ$ since the coverage is incomplete beyond this range. 

We see that the scale length depends weakly on the Galactic azimuth, as shown in Figure \ref{fig:Fig2}. The average scale length determined at various azimuths was found to be 1.95 kpc (marked by a grey dashed line) with a dispersion of 0.26 kpc, shown as $1 \sigma$ shaded region ($1.95 \pm 0.26$) around the dashed line. The values are in agreement with the previous works as indicated in Table \ref{tab:tab1}.

\begin{table*}
    \centering
    \caption{Scale length of the Galactic disc obtained from different studies based on various tracers.\label{tab:tab1}}
\begin{tabular}{|l|l|r|}
\hline
tracers & Scale length (kpc) & reference \\
\hline
\hline
 RC & $1.95 \pm 0.26$ & this study\\
 supergiants & 1.99 $\pm$ 0.13 & \citet{Chrobakova2022}\\
 whole Gaia EDR3 population & 2.19 $\pm$ 0.18 & \citet{Chrobakova2022}\\
 Whole Gaia DR2 population & 2.09 $\pm$ 0.08 & \citet{Chrobakova2020}\\
 OB & 2.10 $\pm$ 0.1 &  \citet{Li2019OB}\\
 RG & 2.13 $\pm$ 0.23 (thin) & \citet{wang2018}\\
 RG & 2.72 $\pm$ 0.57 (thick) & \citet{wang2018}\\
  F and G dwarf & $2.0^{+0.3}_{-0.4}$ (thin) & \citet{Lopez2014} \\
F and G dwarf & $2.5^{+1.2}_{-0.3}$ (thick) &\citet{Lopez2014}   \\
 RC & $2.10^{+0.22}_{-0.17}$ & \citet{lopez2002a}\\
 dust & 2.26 $\pm$ 0.16 & \citet{Drimmel2001}\\
\hline
\end{tabular}

\end{table*}

\subsection{Vertical distribution and Flare modelling}\label{sec:3.2}
\begin{figure}
\includegraphics[width=\columnwidth]{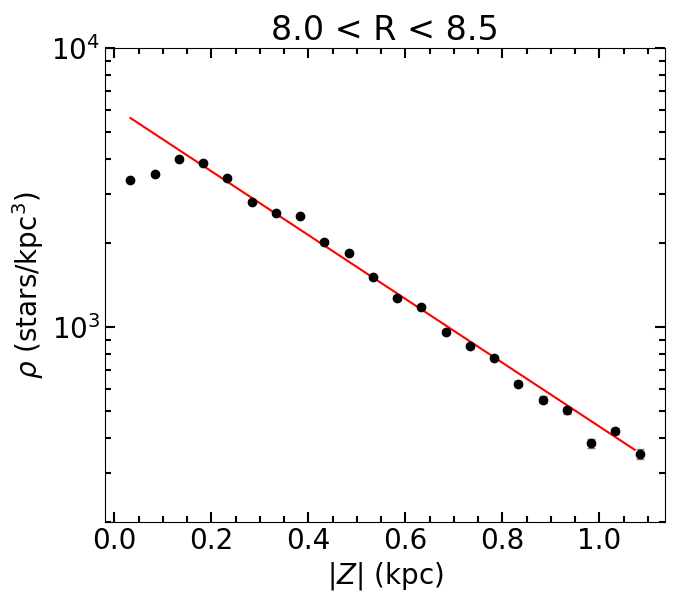}
\caption{Space density of RC stars as a function of absolute height above and below the Galactic plane (Z, in kpc) in the solar neighbourhood, $R = [8.0, 8.5]$ kpc. 1$\sigma$ error bars are included and are smaller than the symbols used. The solid line represents the best-fit model. 
\label{fig:Fig3}}
\end{figure}

\begin{figure}
\includegraphics[width=\columnwidth]{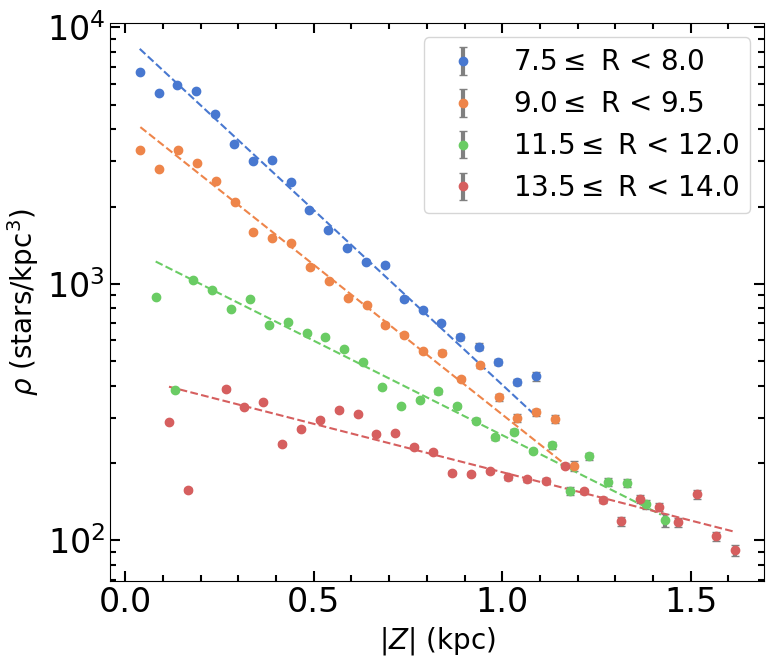}
\caption{Variation of stellar density with $|Z|$ for four bins of Galactocentric distances; $R = [7.5, 8.0]$ kpc in blue, $[9.0, 9.5]$ in orange, $[11.5, 12.0]$ kpc in green, and  [13.5, 14.0] in red. The dashed curves represent the best-fitted model. \label{fig:Fig4}}
\end{figure}

The RC stellar density in the Galactic plane is assumed to decrease in the vertical directions (see the second part of equation \ref{eq:9}). In order to determine the scale height of the old disc, we examined the RC stellar density in off-plane regions. We binned the data into several bins of the Galactocentric radius with a 500 pc bin size.  We used the R-range of $7 \le R \le 16$ kpc and $-50^\circ \le \phi \le +50^\circ$ in our analysis to avoid the low number statistics towards the Galactic centre ($R < 7$ kpc) and contamination in the outer Galaxy (R $>$ 16 kpc). Additionally, the data points closest to the Galactic plane were not included in the analysis to avoid high extinction areas affecting the statistics.  In each R-bin, we examined the change in mean vertical density as a function of height above or below the Galactic plane in 0.05 kpc bins of Z. The variation of stellar density in the solar neighbourhood ($8.0 \le R < 8.5$ kpc) is presented by black points in Figure \ref{fig:Fig3}. It is evident from the figure that the density decreases with height above or below the plane. Equation \ref{eq:9} models the density by assuming the first term, depending on $R$ as approximately constant in $R$ to $R+dR$ bin of Galactocentric radius. We obtained a best-fit scale height of $0.38 \pm 0.1$ kpc in the solar neighbourhood, as represented by the red line in the figure. 
The vertical variation of RC density in various R-bins is analysed, and specific R-bins are highlighted in Figure \ref{fig:Fig4} for illustrative purposes. The RC density is presented for the bins centred at R = 7.75, 9.25, 11.75, and 13.75 kpc in blue, orange, green, and red colours, respectively.  The best-fitted modelled curves are also added to the figure as dashed lines of the same colour. 
 The figure demonstrates that the stellar density at a specific $|Z|$ decreases with increasing Galactocentric radius. Furthermore, the scale height obtained by fitting the data with the exponentially decreasing model in each R-bin is listed in Table \ref{tab:2}.
\begin{table*}
\caption{Scale height of disc in each 0.5 kpc bins of Galactocentric radius obtained by fitting the RC vertical distribution with the z-dependent part of equation \ref{eq:9}.}\label{tab:2}
    \centering
\begin{tabular}{|c|c|c|}
\hline
Radius & Scale height (kpc) &  errors (kpc) \\
\hline
\hline
$7.0 < R < 7.5$ & 0.317 & 0.005\\
$7.5 < R < 8.0$ & 0.318& 0.008\\
$8.0 < R < 8.5$ &  0.380 & 0.009\\
$8.5 < R < 9.0$ &  0.387 & 0.012\\
$9.0 < R < 9.5$ &  0.372 & 0.011\\
$9.5 < R < 10.0$ & 0.371 & 0.011\\
$10.0 < R < 10.5$  & 0.398 & 0.009\\
$10.5 < R < 11.0$  & 0.444  & 0.011\\
$11.0 < R < 11.5$  & 0.495 & 0.013\\
$11.5 < R < 12.0$  & 0.589& 0.020\\
$12.0 < R < 12.5$  & 0.689& 0.026\\
$12.5 < R < 13.0$  & 0.825 & 0.049\\
$13.0 < R < 13.5$  & 0.921 & 0.044\\
$13.5 < R < 14.0$  & 1.151 & 0.085\\
$14.0 < R < 14.5$  & 1.530 & 0.174\\
$14.5 < R < 15.0$  & 1.720 & 0.181\\
$15.0 < R < 15.5$  & 1.911 & 0.249\\
$15.5 < R < 16.0$  & 2.197 & 0.434\\
\hline
\end{tabular}
\end{table*}

\begin{figure}
\includegraphics[width=\columnwidth]{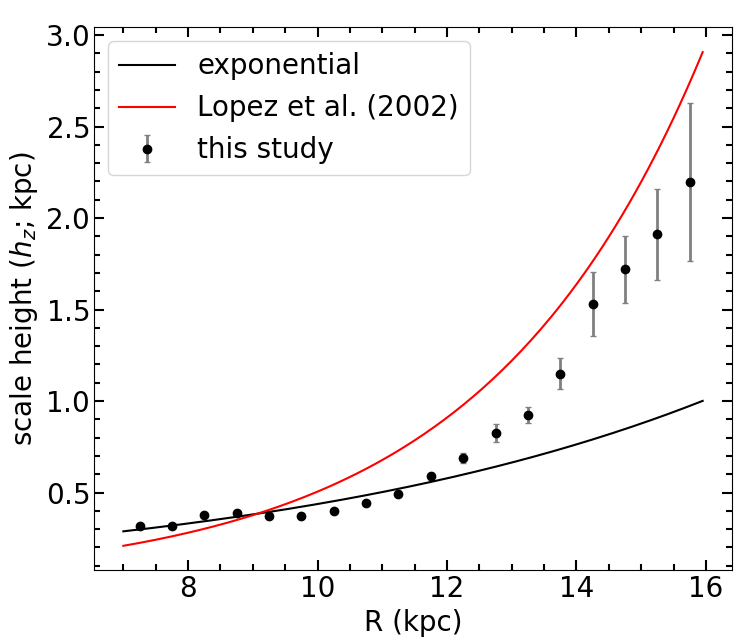}
\caption{Variation of scale height ($h_z$) as a function of Galactocentric distance (R) exhibiting flaring disc traced by RC stars in black points. The black line represents the best-fitted exponential model while the red line denotes the exponential flare from \citet{lopez2002a}. \label{fig:Fig5}}
\end{figure}

Table \ref{tab:2} and black points in Figure \ref{fig:Fig5} show that the scale height increases from 
 $0.380 \pm 0.009$ in the Solar neighbourhood (R = [8.0, 8.5]) to $2.2$ kpc at $\sim 15.75$ kpc. This increasing trend clearly indicates the flaring of the disc, which has also been observed in other tracers. The flaring of the disc using RC stars was also
analysed by \citet{lopez2002a}, but considering only a few lines of sight. In order to compare the flare parameters, we fitted an exponentially increasing scale height model (equation \ref{eq:11}) on our RC sample as described in \citet{lopez2002a}. 
\begin{equation}\label{eq:11}
    h_z(R) = h_z(R_\odot)~exp \left( \frac{R-R_\odot}{H_{R, flare}} \right)
\end{equation}
Where $h_z(R_\odot)$ is a scale length in the Solar neighbourhood and $H_{R, flare}$ is the scale length of the flare. Fitting the equation \ref{eq:11} on the scale height with Galactocentric distance resulted in the solar neighbourhood scale height of $0.35 \pm 0.01$ kpc and the flare scale length of $7.18 \pm 0.93$ kpc, shown by the black line in Figure \ref{fig:Fig5}. The obtained parameters from our study (black line in Figure \ref{fig:Fig5}) differ 
from 
that of \citet{lopez2002a} (represented by red line), where  $h_z(R_\odot) = 0.31^{+0.06}_{-0.04}$ kpc and flare scale length $H_{R, Flare} = 3.4 \pm 0.4$. However, it should be noted that when considering the same volume for analysis, our results show similarity within the error bars.
 
\begin{figure}
\includegraphics[width=\columnwidth]{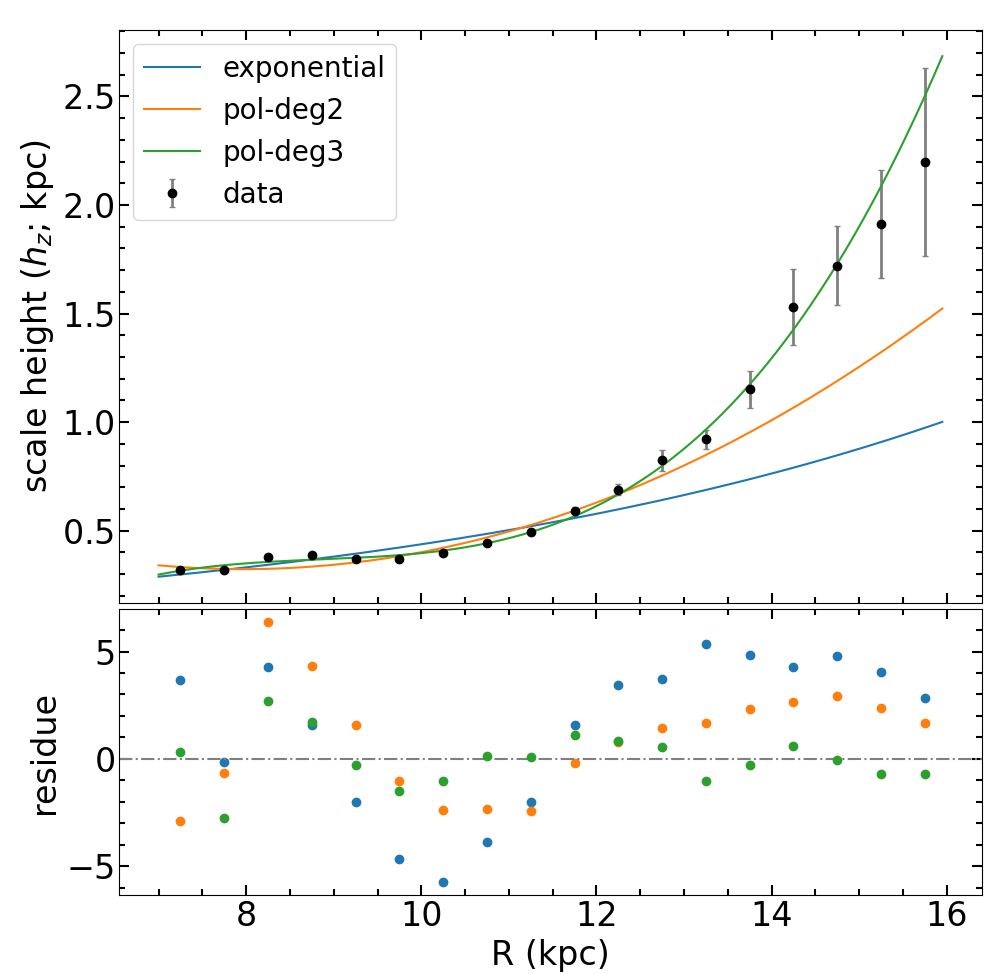}
\caption{Comparison of exponential (in blue), second order polynomial (in orange) and third order polynomial (in green) models of flare fitted on RC scale height data. The weighted residuals corresponding to each model are also plotted in the bottom panel. \label{fig:Fig7}}
\end{figure}

In addition, we noticed that the scale height of RC stars is not best fitted by the exponential model given in equation \ref{eq:11} (see blue line in Figure \ref{fig:Fig7}). The geometric functional form of the flare is not well known; some studies use exponential model \citep{lopez2002a, Li2019OB}, however single order or second order polynomial has also been used in recent times to explain the flare  \citep[][etc.]{Yusifov2004, Chrobakova2022}. We compared exponential (equation \ref{eq:11}) and polynomial models of second (equation \ref{eq:12}) and third order (equation \ref{eq:13}) to understand the geometry of the flare. The fits corresponding to the aforementioned functional forms are plotted in Figure \ref{fig:Fig7} as blue, orange and green curves, respectively. The weighted residual corresponding to each model is also presented in the bottom panel of the figure.  The analysis of the residuals suggests that the third-order polynomial models provide better results compared to the exponential and second-order polynomial models.

\begin{eqnarray} \label{eq:12}
 h_z (R) & = & (0.326 \pm 0.010)~[1 + (0.043 \pm 0.027)~ (R-R_\odot) \nonumber\\& & + (0.058\pm 0.012)~ (R-R_\odot)^2]
\end{eqnarray}
\begin{eqnarray} \label{eq:13}
h_z (R)& =& (0.358 \pm 0.007)~ [1 + (0.060 \pm 0.012)~(R-R_\odot)\nonumber\\ & & +~ (-0.021 \pm 0.011) ~(R-R_\odot)^2 + \nonumber\\ & & (0.017 \pm 0.002)~ (R-R_\odot)^3]
\end{eqnarray}

\subsubsection{Northern and southern flare}\label{sec:3.2.1}  
\begin{figure}
\includegraphics[width=\columnwidth]{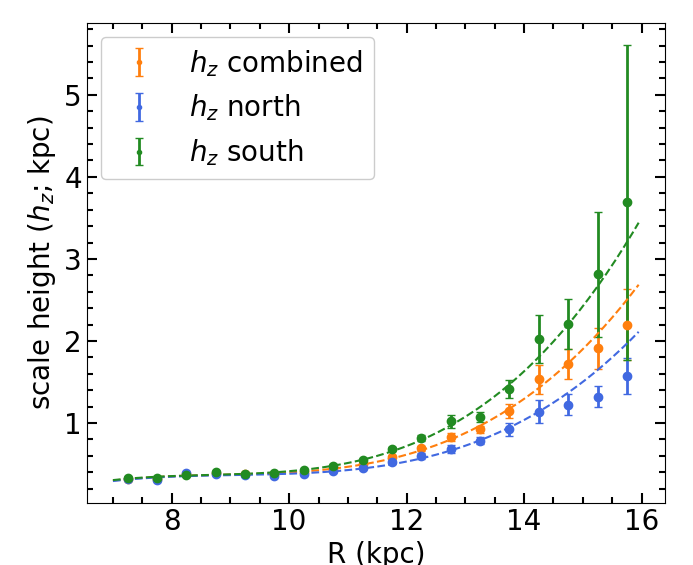}
\caption{Variation of scale height with distance from Galactic centre to compare northern ($Z>0$, in blue), southern ($Z<0$, in green) and combined (north+south, in orange) flare. \label{fig:Fig8}}
\end{figure}

We also explored the asymmetry in the flare with respect to above and below the Galactic plane. In Figure \ref{fig:Fig8}, we plot the scale height as a function of Galactocentric distance for stars above the Galactic plane (in blue), below the Galactic plane (in green) and combined scale height (orange). We notice that there is no significant difference in the scale height of three different sets close to the solar neighbourhood. However, the flaring disc may seem to be asymmetric towards larger distances (R $>$ 11 kpc).
Fitting third-order polynomial on the three cases resulted in the best-fitted parameters listed in Table \ref{tab:tab3}. Analysing the fitted parameters for the Northern and Southern flares reveals that the scale height of the Galactic disc below the Galactic plane is $\sim 0.92$ kpc higher than that of the Northern flare at Galactocentric distance, $R = 15$ kpc. A recent study by \citet{Chrobakova2022} identified a similar asymmetry in the flare for supergiant stars, where the value of $h_z$ for the Southern flare is approximately 1 kpc higher than that of the Northern flare in the outer regions of the Galaxy ($R \ge 15$ kpc). However, no North-South asymmetry was observed in the flare for a sample of LAMOST OB stars as reported by \citet{Yu2021}.
\begin{table}
\caption{Model ($h_z (R) = h_z(R_\odot)~[1 + k_1 (R-R_\odot) + k_2~(R-R_\odot)^2 + k_3~(R-R_\odot)^3]$) parameters for the northern, southern and combined flare. \label{tab:tab3}}
    \centering
{\setlength{\tabcolsep}{3pt}
\begin{tabular}{|l|c|c|c|c|}
\hline
Region & $h_z(R_\odot$) (kpc) & $k_1$ &  $k_2$  & $k_3$ \\
\hline
\hline
 Combined & 0.358 $\pm$ 0.007 & 0.060 $\pm$ 0.012 & -0.021 $\pm$ 0.011 & 0.017 $\pm$ 0.002 \\
 North & 0.353 $\pm$ 0.009 &  0.060 $\pm$ 0.016  & -0.028 $\pm$ 0.014 & 0.014 $\pm$ 0.003\\
 South & 0.360 $\pm$ 0.007 & 0.057 $\pm$ 0.012 & -0.012 $\pm$ 0.011 & 0.021 $\pm$ 0.002\\
\hline
\end{tabular}
}
\end{table}

\subsection{Warp modelling}\label{sec:3.3}
The variation of the distribution of RC stars as a function of the longitude indicates that the disc of our galaxy
is not flat, but it warps upward in one direction and downwards in the other (Figure 6, paper 1). In this section, we analyse
  this structure in greater detail and model the warp observed from RC stars. To ensure the data reliability, we removed the data in the azimuth range of $90^\circ \le \phi \le 270^\circ$ from our analysis because of the incompleteness in the RC sample towards the inner regions of the Galaxy. Additionally, the RC stars having Galactocentric distance $R < 7$ kpc were excluded due to low number statistics. We calculated elevation above the Galactic plane ($Z_w$) as the highest frequency or mode value of Z in 200 pc bins of X and Y. Following the approach of \citet{Chrobakova2022}, we fitted the estimated $Z_w$ values with the warp model defined in equation \ref{eq:14}. 
\begin{equation}\label{eq:14}
    Z_w = [C_w~ (R-R_w)^{\epsilon_w} \sin(\phi-\phi_w)] ~pc
\end{equation}
Where $C_w$, $R_w$, $\epsilon_w$ and $\phi_w$ are the free parameters characterizing the warp. The warp parameters obtained by fitting equation \ref{eq:14} on the elevation of the Galactic plane are given by
\begin{eqnarray}
C_w & = & (0.0057 \pm 0.0050) \\[5pt]
\epsilon_w & = & 1.40 \pm 0.09 \\[5pt]
 \phi_w & = & -2^\circ.03 \pm 0^\circ.18\\[5pt]
 R_w &= &7.4 \pm 0.4\;\; (kpc) 
\end{eqnarray}
    
\begin{figure}
\includegraphics[width=\columnwidth]{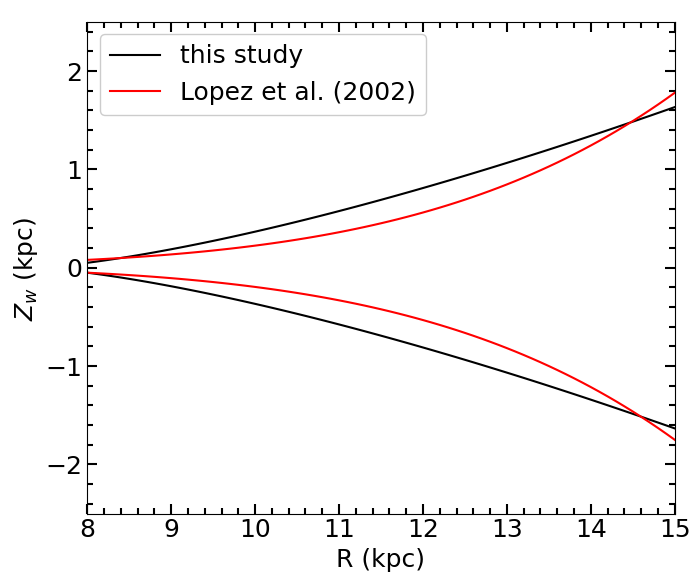}
\caption{Comparison of minimum and maximum modelled warp amplitude for RC stars from our study in black curves with that of \citet{lopez2002a} shown in red lines. \label{fig:Fig9}}
\end{figure}

The maximum and minimum warp amplitude obtained from best-fit model are shown as black curves in Figure \ref{fig:Fig9} and compared with the warp obtained in \citet{lopez2002a} ($C_w = 2.1 \times 10^{-19}$, $\epsilon_w = 5.25 \pm 0.5$, and $\phi_w = -5^\circ \pm 5^\circ$, in red line). 

\begin{figure}
\includegraphics[width=\columnwidth]{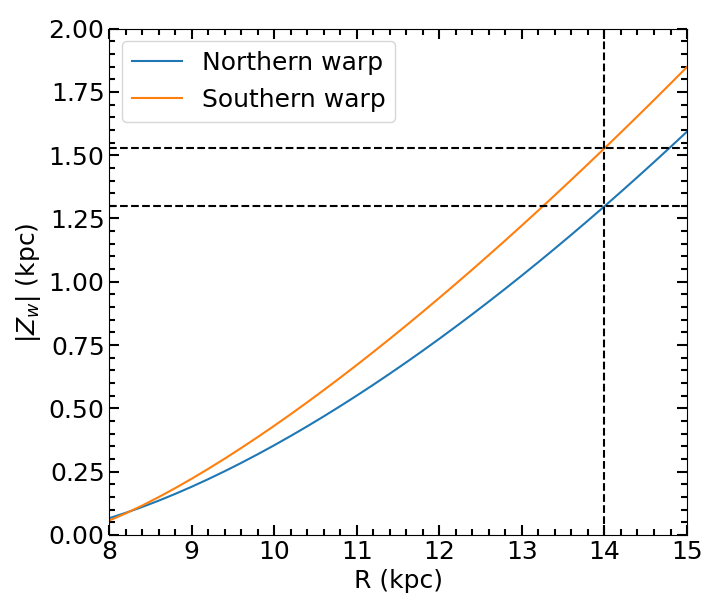}
\caption{A comparison of minimum and maximum warp amplitude for RC stars as a function of radial distance, illustrating the northern and southern warp in blue and orange colours, respectively.\label{fig:Fig10}} 
\end{figure}

Furthermore, we conducted a similar analysis to investigate the north-south asymmetry in the structure of the warp. This was accomplished by separately fitting the warp model (equation \ref{eq:14}) to the northern ($\phi > -2.03$) and southern side ($\phi < -2.03$) of the warp, which was defined based on the nodal line. The maximum amplitude of the northern warp at R = 14 kpc is found to be 1.30 kpc while the minimum amplitude on the southern side is -1.52 kpc, indicating a small asymmetry, as shown in Figure \ref{fig:Fig10}. The asymmetry in the warp is also noted in various studies, a comparison of which is presented in table \ref{tab:tab4}. Notably, 
most studies suggest that the amplitude of the northern warp is greater than the southern. However, our study indicates an opposite trend, which was also observed in the warp traced by RGB stars in \citet{Romero2019} and Cepheids in \citet{Lemasle2022}.  \citet{Chrobakova2022} investigated the whole Gaia EDR3 population, corresponding to an average age of $\sim 6-7$ Gyr and supergiant stars  representing the younger population. The warp amplitude of both the population shows asymmetry, with higher amplitude on the southern side (Gaia EDR3: -0.375; Supergiants: -0.717 ) as compared to the northern (Gaia EDR3: 0.360; supergiants: 0.658) at R = [19.5, 20.0] kpc. 

\begin{table*}
\caption{Values of the warp up ($Z_{up}$) and down ($Z_{down}$) amplitude with respect to the Galactic plane at Galactocentric radius R = 14 kpc expressed in kpc for different tracers. \label{tab:tab4}}
    \centering
\begin{tabular}{|l|c|c|r|}
\hline
Tracer & $Z_{up}$ (kpc) &  $Z_{down}$ (kpc) & references \\
\hline
\hline
RC & 1.30 & -1.52 & this study \\
Cepheid & 0.53 & -0.79 & \citet{Lemasle2022}\\
Cepheids & 0.65 & -0.54 & \citet{Skowron2019b}\\
RGB & 0.97 & -1.22 & \citet{Romero2019}\\
OB & 0.23 & -0.19 & \citet{Romero2019}\\
2MASS NIR & 0.50 & & \citet{Reyle2009}\\
HI & 0.74 & -0.75 & \citet{Levine2006a}\\
dust & 0.74 & -0.68 & \citet{Marshall2006}\\
Pulsar & 0.62 & -0.58 & \citet{Yusifov2004}\\
RC/ 2MASS NIR & 1.23 & - & \citet{lopez2002a}\\
COBE/DIRBE, NIR \& FIR & 1.34 & - & \citet{Drimmel2001}\\
\hline
\end{tabular}
\end{table*}

\section{Discussion}\label{sec:4}
In this work, we have modelled the Milky Way disc using the large sample of RC stars covering a wide area of the Galactic plane. Our investigation of the distribution of RC stars within the disc has revealed that the thickness of the disc is not constant, but it is increasing with the Galactocentric distances, exhibiting a discernible flaring pattern.
In addition, we found that the disc traced by RC stars is not flat but curved in outer directions, indicating a distinct warp feature. We modelled the outer Galactic disc from RC stars and constrained the geometry of flare and warp.
Warp and flare in the Milky Way disc from RC stars have also been discussed in \citet{lopez2002a} (LC02). However, the analyses presented in the previous sections yielded more constrained results than that of LC02. The difference in the obtained parameters can be attributed to two main factors. Firstly, the estimated structure in LC02 is a result of the distribution of RC stars in a few selected 
lines of sight (12 fields, see Table 1 from LC02)
, whereas we have used  RC stars covering nearly the entire Galactic disc in the range $40^\circ \le \ell \le 340^\circ$ and $-10^\circ \le b \le 10^\circ$. 
Secondly, the density model fitting in LC02 suffered from low number statistics, the effect of which is reflected in the large uncertainties in their fitted parameters. These factors may hinder the interpretation of the flare and warp of the disc. 

In addition to LC02, we compare our results with the earlier studies based on different age tracers with the aim of investigating the age dependency on the flare and warp of the disc. 
In the following subsections, we present the comparison of the flare and warp model from RC stars based on our study with that of various tracers from the literature.  We also discuss their implication on the formation mechanisms for the flare and warp. 
\subsection{Age dependency of the Flare}\label{sec:4.1}
\begin{figure}
\includegraphics[width=\columnwidth]{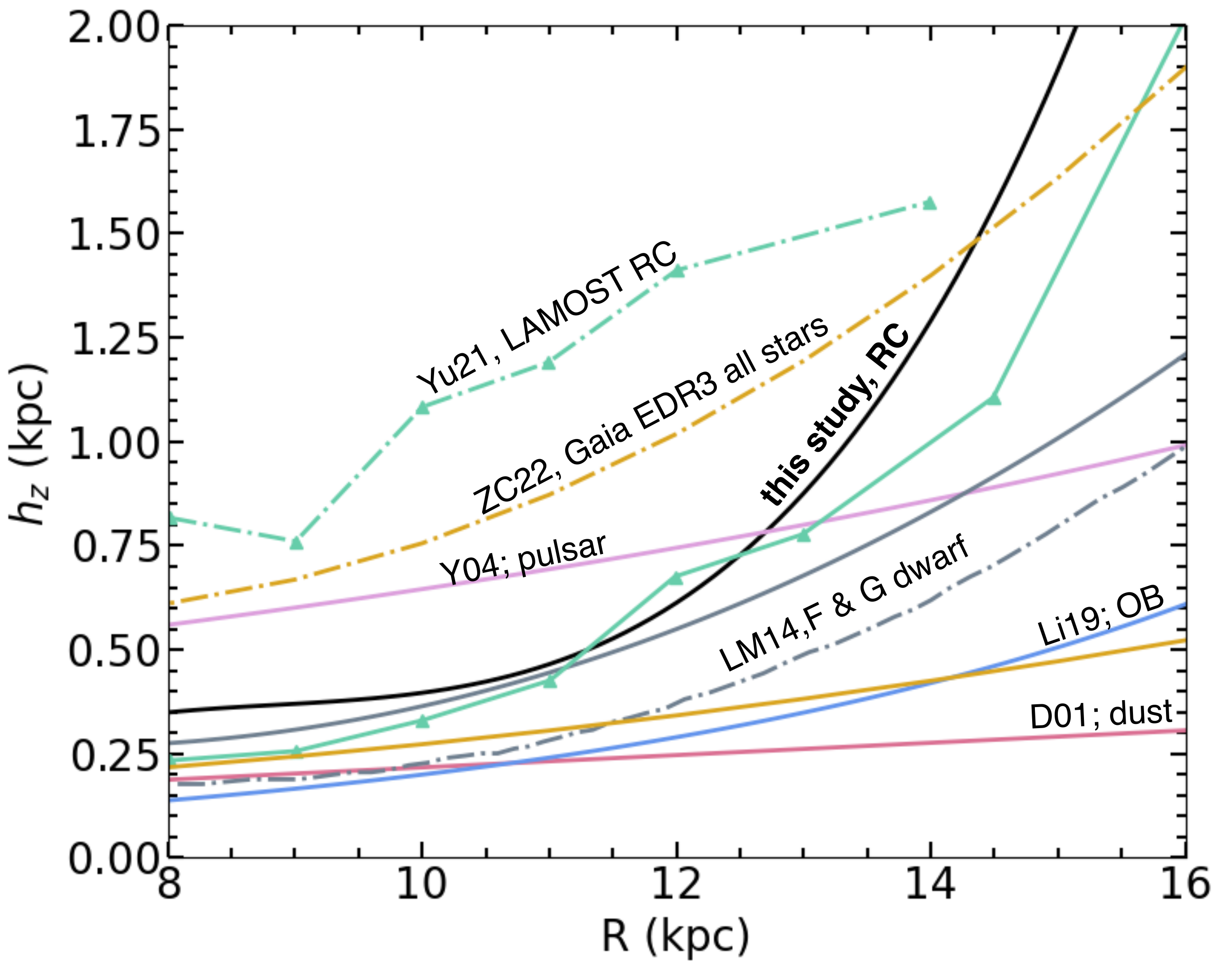}
\caption{Variation of scale height with Galactocentric distance depicting the comparison of disc flare of our study (black line) with the other tracers shown by different colours. The dashed and solid lines of the same colour represent the thick and thin disc of the corresponding tracer. \label{fig:Fig11}}
\end{figure}

Figure \ref{fig:Fig11} displays a comparison of the disc flare revealed from the RC stars in our study (shown in black) with that of previous research, represented by different coloured 
curves. The flare model considered in the figure represents the most recent literature on different tracers, including dust \citep[red,][D01]{Drimmel2000}, pulsar \citep[purple,][Y04]{Yusifov2004}, F and G- type dwarf \citep[grey,][LM14]{lopez2014flare}, OB stars \citep[blue,][Li19]{Li2019OB}, RC stars \citep[light green,][Yu21]{Yu2021}, and whole Gaia EDR3 population \citep[golden,][ZC22]{Chrobakova2022}. The solid and dashed line of the same colour represents the variation of scale height of the thick disc (solid) and thin disc (dashed-dotted) of the corresponding tracer with Galactocentric distance. 
The figure demonstrates that the scale height of the disc increases with Galactocentric radius for all tracers, implying that flaring is a global property of the disc. The phenomenon of flaring in discs is believed to be associated with the process of disc heating, and numerous theories have been proposed to elucidate this phenomenon \citep[e.g.,][]{Cheng2019, Yu2021}. Some internal disturbances such as spiral arms \citep{Sellwood2013}, Giant molecular clouds \citep{Lacey1984} or stellar migration \citep{Bovy2016} have been suggested as potential contributors to the disc flaring events. In addition, external perturbations like the mergers with dwarf or satellite galaxies 
\citep{Kazantzidis2008, House2011} can also play a significant role in heating the disc and causing flaring in the outer regions of the galaxies. The formation mechanisms of a flare can be broadly classified into two types: secular evolution \citep{Narayan2002, Minchev2012, Minchev2015} and the cumulative effect of interaction with passing dwarf galaxies \citep{Kazantzidis2008, villalobos2008, Laporte2018}. Observationally, 
both scenarios expect that the scale height of the disc should increase smoothly with the radius. In the case of secular evolution, the older population is expected to exhibit a stronger flare compared to the younger population. However, no age dependency in flare strength is expected for flares formed by perturbations, as all the stellar populations would be equally disturbed.

Figure \ref{fig:Fig11} presents compelling evidence for the influence of age on scale height as a function of radial distance. Specifically, our findings demonstrate that the older age populations, such as the RC \citep[indicated by the black lines in this study and sea-green line from][]{Yu2021} and all Gaia EDR3 stars (6-7 Gyr average age, in golden colour), exhibits the most pronounced flare. While the flare represented by F and G type dwarf (dashed and dotted grey lines) from LM14 exhibits a less steep curve. Moreover, the scale height of the younger OB stars (in light blue) and the dust (in light red colour) is further lower than the other tracers. Comparing all the curves, we observe a positive correlation between the age of the stars and flare strength.
An age dependency has also been observed in the flare properties of many external galaxies \citep{Streich2016}. This aspect favours the secular evolution of the disc.  
However, a case against secular evolution was presented in \citet{Yu2021} by comparing the flare of OB stars from the LAMOST survey with the RGB population from \citet{wang2018}. Their results indicate almost similar flare strength in younger(OB stars) and older (RGB)  populations, favouring the perturbations caused by the external mergers as the major contributor to the disc heating or formation of flare. Similar results were also suggested in the earlier studies; for example, \citet{wang2018, Xu2020}. Nevertheless, many competing studies \citep{Bovy2012, Li2019OB} still favour the secular evolution of the flare in the disc. 
It has been shown in cosmological simulations that a strong flare can be produced with galaxy merger events and may have an age dependency on scale height \citep{Martig2014}.
In our study, we are not ruling out the possibility of disc perturbation events. This would require additional information from chemo-dynamical studies on a large scale and is out of the scope of the current data set.

\subsection{Age dependency of the Warp}\label{sec:4.2}
In recent years, many models have been proposed to understand the formation of the disc warp (more details in section \ref{Sec:1}). The gravitational origin suggests that all the tracers should exhibit similar warp amplitude. Hence, no age dependency is expected in the case of gravitational origin. However, for the non-gravitational origin, like in-falling gas on the outer parts of the Galaxy, the younger population will get more affected than older ones. The reason is that the young population following the gas distribution will always have a larger warp amplitude, whereas the old population had more time to reduce the amplitude of the warp due to the self-gravity in the models in which the torque affects mainly the gas and not the stars.

\begin{figure}

\includegraphics[width=\columnwidth]{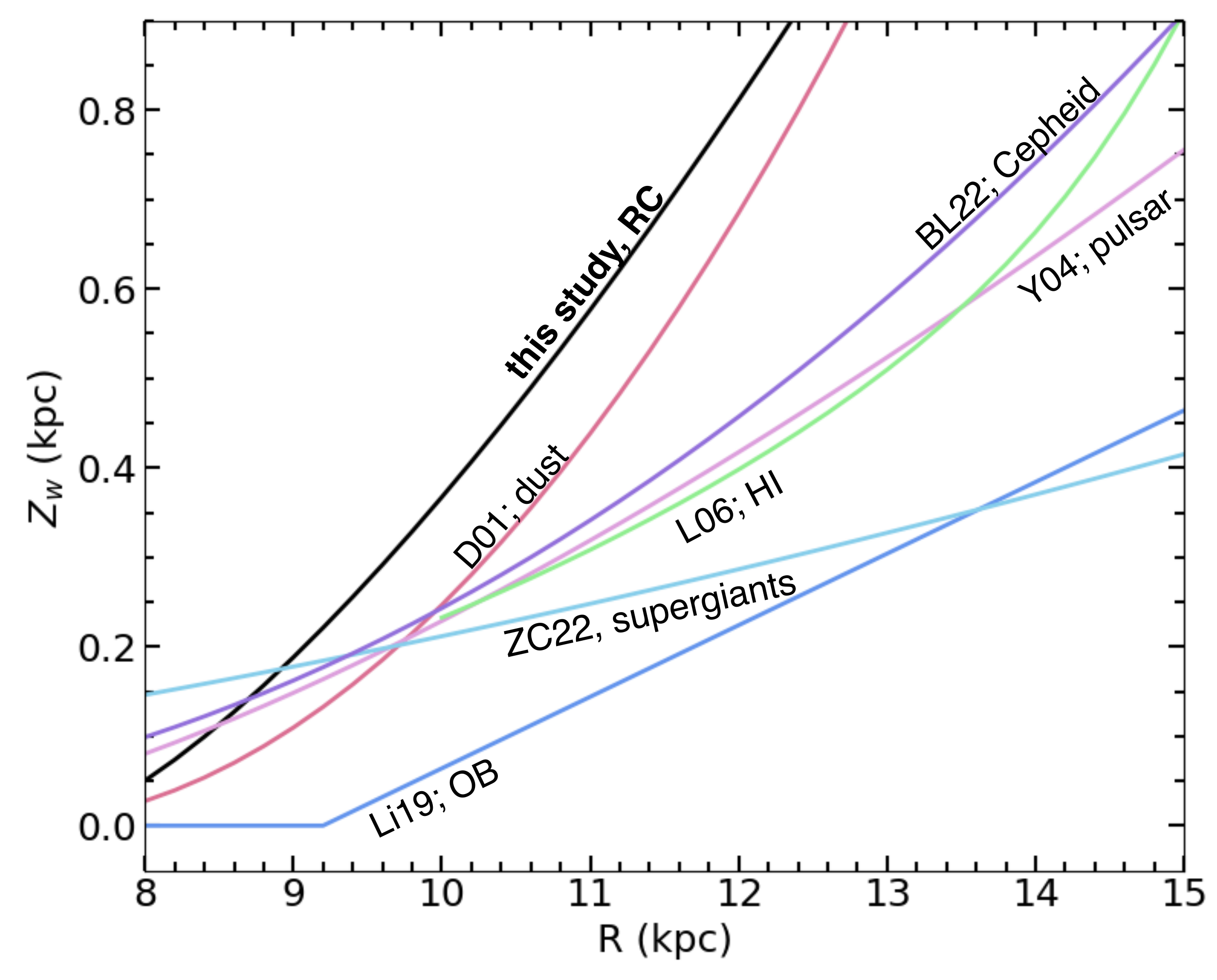}
\caption{Variation of maximum warp height above the Galactic plane with Galactocentric radius for various age tracers outlined by different colours. The result from our study is shown in the black line. \label{fig:Fig12}}
\end{figure}

In Figure \ref{fig:Fig12}, the warp model obtained from RC star distribution in this study (black curve) is presented together with some former works. The figure includes the warp model based on dust \citep[light-red,][D01]{Drimmel2001}, pulsar \citep[purple,][Y04]{Yusifov2004}, HI \citep[light-green,][L06]{Levine2006a}, OB stars \citep[light-blue,][Li19]{Li2019OB}, Cepheid \citep[violet,][BL22]{Lemasle2022}, and supergiants \citep[cyan,][ZC22]{Chrobakova2022}. The analysis conducted by \citet{Chrobakova2020} on a sample of younger stars with $M_G < -2$ mag from Gaia DR2 data and the study of the entire Gaia EDR3 population \citep{Chrobakova2022} also reveals the presence of disc warping. However, the warp amplitudes observed in these populations are not depicted in Figure \ref{fig:Fig12} due to the possibility of mixed populations. Nonetheless, these findings are discussed separately in Appendix \ref{Ap1}.

The different shapes of curves in Figure \ref{fig:Fig12} represent differences in the models used in the respective studies. It is evident from the figure that the warping amplitude varies with the age of the tracer. The difference in the warp amplitude with age implies that the warp is a long-lived feature that is evolving with time, contrary to the idea of a steady warp formed due to gravitational forces alone. Therefore, it must have originated from some non-gravitational effects. The asymmetry in the amplitude of the northern and southern warp, as discussed in section \ref{sec:3.3}, also favours the non-gravitational origin of the warp formation. 

A closer inspection of Figure \ref{fig:Fig12} reveals that the warp amplitude is significantly higher for dust (in green line) and older population (RC in black line) as compared to the younger populations such as Cepheids, pulsar, Supergiant and OB stars. The younger sample of Gaia DR2 (pink curve in Fig \ref{fig:FigA1}) also shows lesser warp amplitude than the whole Gaia EDR3 sample (golden curve), dominated by older populations. This suggests that the warp is more prominent in older tracers relative to the younger ones. This trend aligns with previous studies such as  \citet{warpAmores2017} and \citet{Romero2019}. In contrast, some studies \citep[e.g.,][]{Chen2019,Wang2020,Chrobakova2020, warpLi2023} have reported a decrease in the amplitude of warp with age and support the models in which gas is necessary for warp formation. The varying interpretations of the warp amplitude with age may be attributed to differences in geometric models as well as the region of the Galactic disc covered in different studies. Therefore, identifying the possible mechanism for the formation of the warp is challenging. However, the observed increase in warp amplitude with age could be related to warp dynamics, such as the precession of the 
disc \citep[][]{poggio2020Nat} or accretion of older stars from an interacting galaxy.

\section{Conclusion}\label{sec:5}
In this study, we used our sample of 8.8 million RC stars extracted from 2MASS data in paper 1 to explore the flare and warp in the Galactic disc. We modelled the RC density distribution by assuming exponentially decreasing space density models as a function of Galactocentric radius and height above and below the Galactic plane. Additional terms were added to the model to incorporate the warp and flare of the disc. 
The model-fitted results for the RC stars were compared to the other tracers, and the following conclusions were arrived at:
\begin{enumerate}
    \item The scale length of the old stellar disc weakly depends on the azimuth, and the average scale length is found to be $1.95 \pm 0.26$ kpc. 
    \item The vertical density of RC stars decreases with the height above and below the Galactic plane. The varying scale height with Galactocentric distance exhibits the flaring of the disc. 
    \item The geometry of the flare, as seen in the RC stars, does not follow an exponential increase. It is rather best modelled by a quadratic or cubic polynomial.
    \item Our investigation revealed a slight asymmetry in the scale height of the disc as traced by RC stars, situated both above and below the Galactic plane. We observed a scale height difference of $\sim 1$ kpc at R = 15 kpc, with the southern flare exhibiting a higher scale height compared to the northern counterpart.
    \item The RC distribution in the disc shows that the disc of our galaxy is not flat, but it is warped upward in one direction and downwards in the other. We fit the stellar distribution in the disc using a warp model, $Z_w = C_w (R-R_w)^{\epsilon_w} \sin(\phi - \phi_w)$ with 
    $C_w = 0.0057 \pm 0.0050$, $\epsilon_w = 1.40 \pm 0.09$,  $\phi_w = -2^\circ.03 \pm 0^\circ.18$ and $R_w = 7.4 \pm 0.4$ kpc.
    \item From our analysis, we find that the maximum amplitude of the southern warp, -1.52 kpc, is more than the maximum amplitude of the northern warp (1.30 kpc), showing north-south asymmetry in the warp. 
  
    \item Comparing the flare and warp parameters from the RC distribution to those from different age tracers, we observe a positive correlation both in the flare strength and warp height with the age of the population, providing constraints on  Galaxy evolution models.    
\end{enumerate}

Overall, our study of RC stars covering a large part of the Galactic disc highlights the importance of the requirement of a study of different age populations over a wide area of the Galactic disc. This approach is essential for effectively constraining the formation mechanism of the flare and warp.  
In the future, we hope to combine kinematic  
 information with the stellar distribution of various age populations to investigate these intriguing questions in greater detail. 

\section*{Acknowledgements}
Work at PRL is supported by the Department of Space, Government of India.  We thank the reviewer for detailed comments and suggestions which have improved this work.
We want to thank Dr Chrob{\'a}kov{\'a}, Instituto de Astrofísica de Canarias (IAC), for valuable discussions.  NU and SG thank the Observatoire de la Côte d'Azur for their hospitality during their stay in Nice, France, where some of this work was finalized.  SG also acknowledges a fellowship under the {\it Short Research Trip to France} scheme awarded by the Institut Français, India, which supported his trip.

 This work presents the results from the Two Micron All Sky Survey (2MASS).  2MASS is a joint project of the University of Massachusetts and the Infrared Processing and Analysis Centre/California Institute of Technology, funded by the National Aeronautics and Space Administration and the National Science Foundation. This publication makes use of data products from the European Space Agency (ESA) mission Gaia \href{https://www.cosmos.esa.int/gaia}{(https://www.cosmos.esa.int/gaia)}, processed by the Gaia Data Processing and Analysis Consortium (DPAC; \href{https://www.cosmos.esa.int/web/gaia/dpac/consortium}{https://www.cosmos.esa.int/web/gaia/dpac/consortium}). Funding for DPAC has been provided by national institutions, in particular, the institutions participating in the Gaia Multilateral Agreement.  

\section*{Data Availability}

 The data underlying the research presented in this article is available on request from the authors.



\bibliographystyle{mnras}
\bibliography{ref.bib} 




\appendix

\section{Variation of Warp amplitude with age }\label{Ap1}

\begin{figure}
    \centering
\includegraphics[width=\columnwidth]{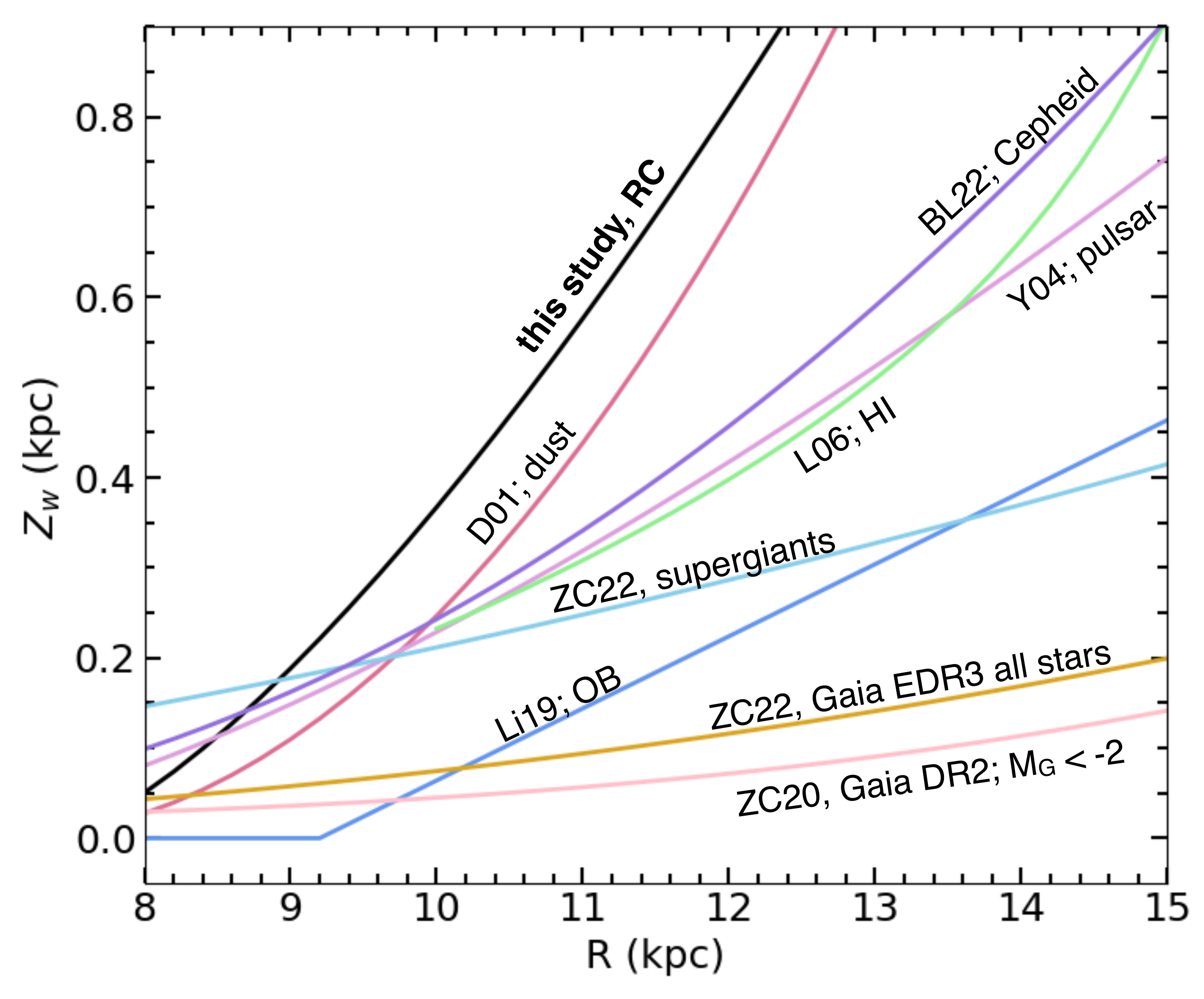}
\caption{Same as Figure \ref{fig:Fig12} but including the warp observed in whole Gaia EDR3 data  (golden curve) and bright stars ($M_G < -2$ mag) from Gaia DR2 data in pink colour. \label{fig:FigA1}}
\end{figure}

This section discusses the variation of warp amplitudes mapped by different age tracers.  Figure \ref{fig:FigA1} showcases the maximum amplitude of the Milky Way disc as derived from our study (in black) and for various other populations from the literature.  It is similar to Figure \ref{fig:Fig12}, with the addition of two new curves representing the warp traced from the  Gaia DR2 bright stars ($M_G < -2$) in pink and from the whole Gaia EDR3 population in golden colour.
It is important to note that the sample of bright stars from Gaia DR2 and all EDR3 stars comprises a heterogeneous mixture of various age populations rather than representing a specific age group. Consequently, comparing the warp traced by these populations with other age tracers is not straightforward. However, we can make a meaningful comparison between the Gaia EDR3 and Gaia DR2 populations. The bright subsample of Gaia DR2 is expected to be younger than the whole population of Gaia EDR3. By comparing the warps traced by these two populations, we observe that the amplitude of the warp increases with the age of the considered population. This observation and its interpretation confirm the age dependence on the Milky Way warp as discussed in Section \ref{sec:4.2}


\bsp	
\label{lastpage}
\end{document}